\documentstyle[preprint,aps,epsfig]{revtex}
\tightenlines
\begin{document}
\draft

\title{Meson PVV Interactions are determined by Quark Loops}
\author{R.~Delbourgo\cite{A1}, Dongsheng~Liu\cite{A2}
 and M.D.~Scadron\cite{A3}}
\address{University of Tasmania, GPO Box 252-21, Hobart,\\
Australia 7001}
\date{\today }
\maketitle

\begin{abstract}
We show that all abnormal parity three-body meson interactions can be
adequately described by quark loops, evaluated at zero external momentum,
with couplings determined by $U(N_f)$ symmetry. We focus primarily on
radiative meson decays which involve one pseudoscalar. The agreement with
experiment for non-rare decays is surprisingly good and requires very few
parameters, namely the coupling constants $g_{\pi qq}$ and $g_{\rho qq}$
and some mixing angles. This agreement extends to some three-body decays
that are dominated by pion pairs in a P-wave state.
\end{abstract}


\narrowtext

\section{INTRODUCTION}
One can perfectly well view $\pi^{o} \rightarrow \gamma\gamma$ decay as 
arising from the $PVV$ quark-level \cite{LSM} linear $\sigma$ model
(L$\sigma$M) instead of through the $AVV$ chiral anomaly \cite{CA}. In such
a description it is straightforward to predict many other $PVV$ radiative
decays through quark triangle graphs. The purpose of this paper is to
investigate how well this quark loop picture accords with experimental
data \cite{PDG}; in this way we avoid discussing anomaly predictions of
these processes and notions of strong CP violation with consequential (low
mass) axions, since these effects have yet to be discovered in the
laboratory. We also consider the situation for rare $PVV$ decays that are
associated with tiny quark mixings (such as $J/\psi\rightarrow \pi\rho,\quad
\eta'\phi$) and with isospin-changing corrections (such as $J/\psi\rightarrow
\pi\omega,\quad\eta\rho$).

In Section II we show that the dominant L$\sigma$M quark triangle $PVV$
predictions for $P\rightarrow \gamma \gamma$, $P\rightarrow V\gamma$ and
$V\rightarrow P\gamma$ all fit the data, within experimental errors; the
only parameters at our disposal are $g_{Pqq}$ and $g_{Vqq}$, the couplings
of the pseudoscalar and vector mesons respectively to the quarks, the
constituent quark masses $m_q$ and a couple of well-known mixing angles.
We pursue this approach in Section III, for the rarer isospin-conserving
interactions where the data gives valuable information about the small amount
of admixture of light quarks in the heavy $q\bar{q}$ states, and vice versa;
here the agreement is less good, but the experimental information is also
sparser and may be expected to change with time. In section IV, we show
that a simple picture of $\Delta I=1$ meson-meson transitions, associated
with the $u-d$ quark mass difference, nicely provides the rates for isospin
changing decays. The last section contains some preliminary work on
three-body decays in which at last one pair of pions is produced in a
P-state; we show that $\rho$-meson poles essentially saturate the amplitudes
for those processes, given the two-body amplitudes determined previously.
The reasonable agreement between naive theory and experiment suggests that
this quark loop picture is a viable alternative to the field-theoretic
anomaly-inspired evaluation of such amplitudes. However the tantalizing
picture which emerges is that internal mass dependence of the box diagrams
is largely negated by the form factors which must be present at the
meson-quark vertices. Such are our conclusions in Section VI.

\section{QUARK TRIANGLE GRAPHS FOR $PVV$ INTERACTIONS}

For the past thirty years, the $\pi^{o}\rightarrow\gamma\gamma$ decay has been
understood as an $AVV$ chiral anomaly \cite{CA} combined with the partial 
conservation of axial current (PCAC). Writing this process and all other $PVV$
amplitudes in the conventional form $M_{PVV}\epsilon_{\mu\nu\kappa\lambda}
e'^\mu k'^\nu e^\kappa k^\lambda$, where $k$ and $e$ refer to vector particle
momentum and polarization respectively, one derives
\begin{equation}
M_{\pi^{o}\gamma\gamma} = \frac{e^{2}N_c} {12\pi^{2}f_{\pi}} = 
\frac{\alpha N_c}{3\pi f_{\pi}} \approx 0.025 {\rm~GeV}^{-1}. 
\end{equation}
With $f_{\pi} \approx 93$ MeV and a quark color number $N_c = 3$,
the theoretical result is accurately consistent with data \cite{PDG}:
\begin{equation}
\mid M^{exp}_{\pi^{o}\gamma \gamma} \mid = \sqrt{64\pi
\Gamma_{\pi^0\gamma\gamma}/m^{3}_{\pi}}
= 0.025 \pm 0.001 {\rm~GeV}^{-1}.
\end{equation}
The question then arises as to which theoretical models also predict the same 
result while simultaneously demanding that $N_{c}=3$.

One such favored model is the quark-level \cite{LSM} linear $\sigma$ model
(L$\sigma$M). This model also provides $u$ and $d$ quark triangle 
graphs but with the meson interacting through the pseudoscalar $\gamma_5$
vertex, rather than through the divergence of an axial vector. No accompanying
L$\sigma$M $\pi^{+}$ loop is allowed because the $\pi^{o}\pi^+\pi^-$
coupling vanishes; consequently the $AVV$ anomaly result immediately follows
in the spirit of Steinberger \cite{Steinberger}. (In carrying out the
evaluation of $M$, we go to the soft limit, because the pion is essentially
massless, relative to other hadronic masses and even the constituent quarks.)
The fact that this L$\sigma$M also demands $N_c=3$ \cite{LSM} is tied to the
Lee null tadpole condition \cite{BWL} and ensures dynamical generation of
masses, in contrast to the original nucleon-level L$\sigma$M \cite{GML}. The
pion-quark coupling $g_{\pi q q}$ needed obeys the quark-level
Goldberger-Treiman relation (GTR) $g_{\pi qq}= m_{ns}/f_{\pi}$ where the
constituent (nonstrange) quark mass $m_{ns}\equiv\hat{m}$ is dynamically
generated to the expected $M_N/3$ value in the chiral limit, since $N_{c}= 3$
is required now.

Our aim in this paper is to investigate how well the sigma model picture fits
the data for $PVV$ processes. We take the naive viewpoint that the $PVV$
amplitudes, just like $\pi^0\gamma\gamma$ can be estimated by working out
$M_{PVV}$ in the limit of {\em zero} external momentum, even though this is
a {\em long way} from the mass shell for the {\em heavy} mesons particularly.
The reason why we do so is because many of these processes are quite close
to particle thresholds and would otherwise be highly dependent on binding
energies and the like --- for which there is no evidence. In such a soft
limit and including the crossed diagram, it is very simple to work out that
\begin{equation}
 M_{P_1V_2V_3} = \frac{g_{P_1}g_{V_2}g_{V_3}N_c}{2\pi^2} \int
\frac{m_1\alpha+m_2\beta+m_3\gamma}{m_1^2\alpha+m_2^2\beta+m_3^2\gamma}
\delta(1-\alpha-\beta-\gamma)\,d\alpha\,d\beta\,d\gamma,
\end{equation}
where $m_i$ are the internal quark masses, in an obvious cyclic notation. The 
Feynman parametric integral is especially easy to work out when all quark
masses are equal to $m$, when
$$ M_{P_1V_2V_3} \rightarrow \frac{g_{P_1}g_{V_2}g_{V_3}N_c}{4\pi^2m}. $$
It is not much harder to determine the integral (3) if two masses are equal,
say $m_1=m_2=m,\,\, m_3=m'$, when
$$ M_{P_1V_2V_3} \rightarrow \frac{g_{P_1}g_{V_2}g_{V_3}N_c}{2\pi^2}.
 \frac{J(r)}{m+m'}, \quad J(r)\equiv 1 + \frac{2r}{r^2-1}-\frac{4r\ln r}
 {(r^2-1)^2}; \quad r\equiv\frac{m}{m'}. $$
Thus $J(1) = 2$ in the equal mass limit $m = m'$. See Figure 1 to see how 
$J(r)$ varies with $r$.

It only remains to substitute the couplings of the mesons to the quarks in
order to derive the theoretical values for $M_{PVV}$. To do so we will adopt
the naive perspective that all strong couplings are related by $U(N_f)$
symmetry. Therefore, in the absence of any isospin breaking and neglecting
the third generation of quarks, let us see how far one can go with the
effective Lagrangian (suppressing the $\gamma_5$ matrix between spinors),
\begin{eqnarray}
{\cal L}_{Pqq}&=g_P[&\pi^0(\bar{u}u-\bar{d}d)/\sqrt{2}+\pi^+\bar{u}d +
         \pi^-\bar{d}u+\eta_{ns}(\bar{u}u+\bar{d}d)/\sqrt{2}\nonumber\\
& & +K^+\bar{u}s+K^-\bar{s}u+K^0\bar{d}s+\bar{K}^0\bar{s}d
    +\eta_s\bar{s}s+\eta_c\bar{c}c       \nonumber \\
& & +D^0\bar{u}c+D^-\bar{d}c+\bar{D}^0\bar{c}u+D^+\bar{c}d+
     D_s^-\bar{s}c + D_s^+\bar{c}s].
\end{eqnarray}
The universal coupling $g_P$ above is fixed by the $\pi^0$ coupling, which
in turn is determined by GTR; thus $g_P = (m_u+m_d)/\sqrt{2}f_\pi \equiv 
\sqrt{2}\hat{m}/f_\pi \simeq 5.13$. We must also relate the physical neutral
pseudoscalar states to the isospin 0 combinations in Eq. (4) via certain
mixings. For the purpose of this and the next section, we shall disregard
electromagnetic effects in the strong sector and take (see Appendix)
$$\eta_{ns} = \eta\cos\phi_P + \eta'\sin\phi_P;\quad
  \eta_s = -\eta\sin\phi_P + \eta'\cos\phi_P;\quad
  \phi_P\simeq 42^o,$$
with $\eta_c$ unmixed. (Later on we will examine rare decays, where the small 
residual mixings become rather important.)  The strong vector meson
interactions are likewise written as
\begin{eqnarray}
{\cal L}_{Vqq}&=g_V[&\rho^0(\bar{u}u-\bar{d}d)/\sqrt{2}+\rho^+\bar{u}d +
     \rho^-\bar{d}u+\omega_{ns}(\bar{u}u+\bar{d}d)/\sqrt{2}\nonumber\\
& & +K^{*+}\bar{u}s+K^{*-}\bar{s}u+K^{*0}\bar{d}s+\bar{K}^{*0}\bar{s}d
    +\omega_s\bar{s}s+\omega_c\bar{c}c       \nonumber \\
& & +D^{*0}\bar{u}c+D^{*-}\bar{d}c+\bar{D}^{*0}\bar{c}u+D^{*+}\bar{c}d+
     D_s^{*-}\bar{s}c + D_s^{*+}\bar{c}s],
\end{eqnarray}
suppressing vector indices and the $\gamma$ matrix between quark spinors. Here 
the coupling constant is fixed by that of the $\rho$-meson which is itself 
determined by the leptonic decay rate: $g_V = g_\rho/\sqrt{2}\simeq 3.56$.
These vector states also undergo the semistrong mixing via
$$\omega_{ns} = \omega\cos\phi_V + \phi\sin\phi_V;\quad
  \omega_s = -\omega\sin\phi_V + \phi\cos\phi_V;\quad
  \phi_V\simeq 3.8^o.$$
(At this level we may identify $\omega_c$ with the $J/\psi$ particle.)
Finally, it is trivial to consider radiative decays, by substituting $\gamma$
for the meson and including the proper electromagnetic coupling of the
photon to the quarks, $e_q$.

The magnitudes of the $M_{PVV}$ can be found directly from experimental
decay rates, as they occur \cite{PDG}. For $P\rightarrow V V$ decays, the
rate is given by
\begin{equation}
\Gamma_{PVV} = \Delta^3|M_{PVV}|^2/32\pi m_P^3 = p_{VV}^3|M_{PVV}|^2/4\pi,
\end{equation}
where $\Delta = 2m_P p_{VV}$ and $p_{VV}$ is the magnitude of the
three-momentum of one of the vector mesons in the rest frame of the decaying
$P$. (We must be careful to multiply the right-hand-side of (6) by a factor
of 1/2 when the two vector mesons are identical.) A very similar formula
applies to the decay rate $V\rightarrow PV$,
\begin{equation}
\Gamma_{VPV} = \Delta^3|M_{PVV}|^2/96\pi m_V^3 = p_{VP}^3|M_{PVV}|^2/12\pi,
\end{equation}
where $m_V$ refers to the mass of the initial vector and $p_{VP}=\Delta/2m_V$
is the momentum of one final particle in the decay rest frame.
It is then a simple matter to extract the $|M|$ from the measured \cite{PDG}
rates $\Gamma$, for subsequent comparison with theory. We shall do this
constantly, without further elaboration.

We mentioned at the start that the $M_{\pi^0\gamma\gamma}$ amplitude is
theoretically determined in the sigma model by ing
the $u$ and $d$ loop
contributions and equals $e^2g_P/4\sqrt{2}\pi^2\hat{m} = \alpha/\pi f_\pi$,
fitting experiment admirably. Indeed many of the processes involve neutral
mesons and require the interaction terms,
\begin{equation}
g_P[\pi^0(\bar{u}u-\bar{d}d)/\sqrt{2}+(\eta\cos\phi_P+\eta'\sin\phi_P)
 (\bar{u}u+\bar{d}d)/\sqrt{2}
 +(-\eta\sin\phi_P +\eta'\cos\phi_P)\bar{s}s+\eta_c\bar{c}c],  
\end{equation}
\begin{equation}
g_V[\rho^0(\bar{u}u-\bar{d}d)/\sqrt{2}+(\omega\cos\phi_V+\phi\sin\phi_V)
 (\bar{u}u+\bar{d}d)/\sqrt{2}
 +(-\omega\sin\phi_V +\phi\cos\phi_V)\bar{s}s+\psi\bar{c}c]. 
\end{equation}
Instead of treating each process one at a time, let us consider just three
examples to explain what we are doing, before presenting our results for
non-rare decays in a table.

Consider first the decay $\eta' \rightarrow \omega \gamma$, which contains a
charge coupling, one vector coupling and one scalar coupling. The sum of the
loops of nonstrange quarks and the mixing angles implied by (8) and (9)
gives ($e=\sqrt{4\pi\alpha}\simeq 0.3028$)
\begin{equation}
 M_{\eta'\omega\gamma} = eN_cg_Vg_P\sin\phi_P\cos\phi_V/24\pi^2\hat{m}
 \simeq \sqrt{2}eg_V\sin\phi_P/8\pi^2f_\pi.
\end{equation}
Using the values quoted previously, we obtain the rough theoretical magnitude,
$ M_{\eta'\omega\gamma} \simeq 0.139$ GeV$^{-1}$, which can be compared with
the experimental result,
$$ M^{exp}_{\eta'\omega\gamma}= \sqrt{4\pi\Gamma_{\eta'\omega\gamma}/
  p_{\omega\gamma}^3} \simeq 0.137 {\rm~GeV}^{-1}.$$
Our second example is the semistrong process $\phi\rightarrow\pi\rho$, which
is governed by nonstrange quark loops but is rather sensitive to the vector
meson mixing angle $\phi_V$. The theoretical result here is
\begin{equation}
 M_{\phi\rho\pi^0}=\sqrt{2}N_cg_V^2g_P\sin\phi_V/8\pi^2\hat{m}
 \simeq 3g_V^2\sin\phi_V/4\pi^2f_\pi.
\end{equation}
Inserting the accepted value of $\phi_V \simeq 3.8^0$, the prediction is
that $M_{\phi\pi^0\rho}\simeq 0.69$ GeV$^{-1}$, and this is in agreement
with the upper bound obtained by experiment: $M^{exp}_{\phi\pi\rho} < 1.18$
GeV$^{-1}$; in this connection we note that the 1996 PDG compilation 
\cite{1996} cites a
specific value for this particular channel (which we prefer), but this has
been retracted in the 1998 PDG compilation, resulting in less predictive
power. Finally we take a look at the decay $\psi\rightarrow\eta_c\gamma$,
which is governed by a charm loop and involves one photon. This time we get
\begin{equation}
 M_{\psi\eta_c\gamma} = eN_cg_Vg_P/6\pi^2 m_c
 \simeq eg_V\hat{m}/\sqrt{2}\pi^2f_\pi m_c,
\end{equation}
which we estimate to be about 0.19 GeV$^{-1}$, for a constituent quark mass
ratio of $\hat{m}/m_c \simeq 338/1500 \simeq 0.225$. This is close to
average experimental value of $M^{exp}_{\psi\eta_c\gamma} \simeq 0.17$
GeV$^{-1}$, obtained from the decay rate.

In this manner we can work out theoretical estimates of all dominant $PVV$
processes. We have tabulated the results and the experimental magnitudes for
comparison in Table 1. The near agreement of naive theory with data is
actually quite astonishing, because we are dealing with sizeable external
masses in many instances, particularly for heavy charm quarks, and so are a
very long way from the chiral limit. We will return to this point towards
the end of the paper.

It should be noted that the last five decay channels in Table I correspond
to unequal mass quarks running round the loop, in contrast to all the
previous cases, and the absolute value of the amplitudes are perhaps not as
impressively predicted are the others (although the ratio $|M_{K^{*+}K^+
\gamma}/M_{K^{*0}K^0\gamma}|$ is reasonably close to experiment). On the
whole the fit is quite satisfactory, considering the fact that we have no
free parameters at our disposal.

\section{RARE ISOSPIN-CONSERVING DECAYS}

There are many processes involving the heavier quarks which conserve isospin.
They tend to be rarer than the cases considered above and measure the
smaller admixture of lighter quarks in the meson composites. In fact these
rare processes provide much valuable information about such small mixings
and, emboldened by the success of the soft momentum predictions of $PVV$
amplitudes, we shall apply the same method to probe the mixing angles of
neutral vectors and pseudoscalars.  Let us write the meson which couples to
the $\bar{c}c$ as the combination $\eta_c+\delta_c\eta +\delta'_c\eta',$
where the two $\delta$ are very small quantities (which is why we have not
modified the normalization factor in front of $\eta_c$ in first
approximation). It follows then that the non-strange and strange
pseudoscalars are (in the same limit),
\begin{eqnarray} 
\eta_{ns}=&\eta\cos\phi_P + \eta'\sin\phi_P - \eta_c(\delta_c\cos\phi_P +
\delta'_c\sin\phi_P) \nonumber \\
\eta_s =&-\eta\sin\phi_P + \eta'\cos\phi_P + \eta_c(\delta_c\sin\phi_P -
\delta'_c\cos\phi_P).
\end{eqnarray}
Consequently the interactions (8) of the neutral pseudoscalars get modified to
\begin{eqnarray}
&g_P[\pi^0(\bar{u}u - \bar{d}d)/\sqrt{2}+\eta\{\cos\phi_P
(\bar{u}u+\bar{d}d)/\sqrt{2}-\sin\phi_P\bar{s}s+\delta_c\bar{c}c\}\nonumber\\
&+\eta'\{\sin\phi_P(\bar{u}u+\bar{d}d)/\sqrt{2} + \cos\phi_P\bar{s}s 
+ \delta'_c\bar{c}c\} \nonumber\\
&+\eta_c\{-(\delta_c\cos\phi_P+\delta'_c\sin\phi_P)
(\bar{u}u+\bar{d}d)/\sqrt{2}  + (\delta_c\sin\phi_P-\delta'_c\cos\phi_P)
\bar{s}s + \bar{c}c\}].
\end{eqnarray}
Much the same idea can be applied to the neutral vector interactions (with
mixing parameters $\delta$ replaced by $\epsilon$):
\begin{eqnarray}
&g_V[\rho^0(\bar{u}u - \bar{d}d)/\sqrt{2}  +\omega\{\cos\phi_V(\bar{u}u+
 \bar{d}d)/\sqrt{2} - \sin\phi_V\bar{s}s + \epsilon_c\bar{c}c\} \nonumber\\
&+\phi\{\sin\phi_V(\bar{u}u+\bar{d}d)/\sqrt{2} + \cos\phi_V\bar{s}s 
+ \epsilon'_c\bar{c}c\} \nonumber\\
&+\psi\{-(\epsilon_c\cos\phi_V + \epsilon'_c\sin\phi_V)
(\bar{u}u+\bar{d}d)/\sqrt{2}+(\epsilon_c\sin\phi_V -\epsilon'_c\cos\phi_V)
\bar{s}s + \bar{c}c\}].
\end{eqnarray}

We are now in a position to tackle rare $PVV$ processes involving $\eta_c$
and $\psi$. These are evaluated in the same way as before, but now using the
neutral couplings (14) and (15). The results are summarized in Table II;
theoretical values are only worked out to first order in $\delta$ or
$\epsilon$, since these are already small quantities. We need to give some
words of explanation as to what entry in the last column is being predicted
and what is not, in contrast to Table I, which is essentially parameter-free.
In our parametrizations (14) and (15), the cleanest predictor of the amount
of nonstrange and strange quark components in $\eta_c$ are the two decays
$\eta_c\rightarrow 2\rho$ and $\eta_c\rightarrow 2\phi$; they determine
separate combinations of the mixings $\delta_c$ and $\delta'_c$ (which
represent the amount of charm quarks contained in $\eta$ and $\eta'$).
We may derive a reasonable fit by choosing the values $\delta_c\simeq 0.0054$
and $\delta'_c\simeq -0.0008$; this permits us to {\em predict} the decay
rates for $\eta_c\rightarrow 2\omega$ and $\eta_c\rightarrow K^* \bar{K}^*$
in the table. While the latter prediction is good, the former is decidedly
not -- in fact the experimental value is slightly embarrassing and difficult
to understand on any sensible theoretical basis; for U(2) symmetry between
$u$ and $d$ entails that the process $\eta_c\rho^0\rho^0$ should be almost
equal to $\eta_c\omega\omega$, which is a far cry from what seems to be
observed!

With respect to the $\epsilon$ parameters (because $\phi_V$ is so small) the
process $\psi\rightarrow\pi^0\rho^0$ tells us about the nonstrange-charm
mixing $\epsilon_c$ almost at once, while $\psi\rightarrow\eta\phi$ provides
the mixing between strange and charm states, namely $\epsilon'_c$. We find a
fair fit (including sign) with the values, $\epsilon_c\simeq -0.0002,\,
\epsilon'_c\simeq 0.00009$. This enables us to {\em predict} the amplitudes
for the other entries in Table II involving $\epsilon$. They are all roughly
correct, except for $\psi\rightarrow\eta'\gamma$ and $\psi\rightarrow
\eta'\omega$. We do not understand the reasons for this; blaming the
discrepancy on the effect of mass extrapolation from the shell to
zero-momentum would undermine the other reasonable answers; but in any case
it should be noted that the vector meson dominance (VMD) prediction is a
factor of 250 larger than the data which heightens our suspicions. It is
possible that the experimental results may shift a little over time and
bring the results into closer agreement with theoretical expectations, or
possibly the experimental decay rate also includes the contribution from
$a_0(980)$.

\section{Rare $\Delta I=1$ VPV decays}

Here we want to take a look at four processes,
$$\psi\rightarrow\pi^0\omega,\quad \psi\rightarrow\pi^0\phi,\quad \psi
\rightarrow\eta\rho^0, \quad \psi\rightarrow\eta'\rho^0,$$
because they involve isospin breaking interactions of mesons, without being
accompanied by photons. (Interestingly, there seems to be no comparable data
in the $\eta_c$ decay sector.) All four processes are smaller by at least a
factor of three compared to processes that conserve isospin; eg
$M^{exp}_{\psi\pi^0\omega} \simeq 0.3\,M^{exp}_{\psi\pi^0\rho^0}$.

We may ascribe the existence of these small amplitudes to electromagnetic
effects and the $u-d$ mass difference. With regard to the quark triangle
loop, the $u-d$ difference gives a correction of order $\sim 5/350$, which
is insufficient to explain the observed magnitudes. We shall find below that
the experimental results can be reasonably well obtained by $\Delta I=1$
mixings between mesons (including the well-known $\rho-\omega$ transition),
which are themselves dominated by the $u-d$ difference in quark-loop
self-energy contributions.

Before going further, let us estimate two strong $PVV$ amplitudes that
cannot be directly measured from decays, because of the masses of the
participating mesons; they are nonetheless important in a subsidiary role
for what follows. Firstly, there is the interaction $\rho^0\pi^0\omega$
(which actually enters $\omega\rightarrow 3\pi$, via pole dominance), that
we estimate to be
\begin{equation}
 M_{\rho^0\pi^0\omega} = 3\sqrt{2}g_V^2g_P\cos^2\phi_V/8\pi^2\hat{m}\simeq
 3g_\rho^2/8\pi^2f_\pi \equiv 3C \simeq 10.4 {\rm~GeV}^{-1}.
\end{equation}
Next we shall need the amplitude
\begin{equation}
M_{\rho^0\eta\rho^0} = 3C\cos\phi_P \simeq 7.85 {\rm~GeV}^{-1}.
\end{equation}
Finally we shall require amplitudes which we have estimated previously
fairly well (or which we can take directly from experiment), namely
$M_{\psi\pi^0\rho^0} \simeq 0.0021$ GeV$^{-1}$, $M_{\psi\eta\omega}
\simeq 0.0016$ GeV$^{-1}$ and $M_{\phi\pi^0\rho^0}\simeq 0.69$ GeV$^{-1}$.

Other quantities we will need are the $\langle P|H|P' \rangle$ and
$\langle V|H|V'\rangle$ elements, where unprimed states refer to $I=0$
composites while primed states correspond to $I=1$ mesons. These elements
include the most well-studied case of the $\rho-\omega$ transition, which is
the sum of two pieces \cite{CB}:
\begin{eqnarray}
M_{\rho\omega} = \langle \rho|H^{\Delta I=1}|\omega\rangle&=&\langle
\rho|H^{JJ}|\omega\rangle + \langle \rho|H^{u-d}|\omega\rangle\nonumber \\
& \simeq& 700 - 5100 \simeq -4400 {\rm~MeV}^2
\end{eqnarray}
In similar vein, we may estimate the $\rho-\phi$ transition to be
\begin{equation}
M_{\rho\phi} = \langle \rho|H|\phi\rangle \simeq \tan\phi_V
\langle\rho|H|\omega\rangle \simeq -290 {\rm~MeV}^2.
\end{equation}
We must be a little more careful with the $\rho-\psi$ transition, which is
dominated by the photon exchange contribution $H^{JJ}$:
\begin{eqnarray}
M_{\rho\psi} = \langle\rho|H^{\Delta I=1}|\psi\rangle &=&
\langle\rho|H^{JJ}|\psi\rangle + \langle \rho|H^{u-d}|\psi\rangle \nonumber\\
& \simeq & (e m_\rho m_\psi)^2/g_\rho g_\psi + \epsilon_c
\langle\rho|H^{u-d}| \omega\rangle\nonumber \\
& \simeq & 1010 + 1 \simeq 1010 {\rm~MeV}^2.
\end{eqnarray}
Last, but not least, we shall require the $\pi^0-\eta$ and $\pi^0-\eta'$
transitions. This arises from a quark self-energy bubble plus an $a_0$
tadpole term. The latter equals the former contribution \cite{DS} because
$g_{a_0\eta_{NS}\pi} = (m_{a_0}^2-m_{\eta_{ns}}^2)/f_\pi$ by SU(6) and chiral
symmetry \cite{CS}. Noting the gap equation \cite{LSM} stemming from the
GTR, the total result sums to
\begin{eqnarray}
\langle \eta_{ns}|H^{u-d}|\pi^0\rangle &=& 4iN_cg_P^2\int \frac{d^4p}
{(2\pi)^4} [\frac{1}{p^2-m_d^2}-\frac{1}{p^2-m_u^2}]\nonumber \\
&\simeq& 2(m_u^2-m_d^2) \simeq -5400 {\rm~MeV}^2 \quad {\rm for~~}
m_d-m_u\simeq 4 {\rm ~MeV}.
\end{eqnarray}
We thereby estimate the two neutral pseudoscalar elements,
\begin{eqnarray}
M_{\eta\pi} = \langle\eta |H|\pi^0\rangle \simeq -5400\cos\phi_P \simeq -4000
 {\rm~MeV}^2.\nonumber \\
M_{\eta'\pi} = \langle\eta' |H|\pi^0\rangle \simeq -5400\sin\phi_P\simeq
 -3600 {\rm~MeV}^2.
\end{eqnarray}
Actually the last element leads to very small corrections because each of
the mesons is rather far from the mass shell of the other.

We are now in a position to estimate these rare amplitudes. (Remember that
we may ignore the negligible correction induced by $u-d$ mass differences
within the triangle loop.) The first case is
\begin{equation}
M_{\psi\pi^0\omega} = \frac{M_{\psi\rho}M_{\rho^0\pi^0\omega}}{m_\psi^2 -
m_\rho^2} + \frac{M_{\omega\rho}M_{\psi\pi^0\rho^0}}{m_\omega^2-m_\rho^2} +
\frac{M_{\pi\eta}M_{\psi\eta\omega}}{m_\pi^2-m_\eta^2} \simeq 0.0007
{\rm~GeV}^{-1},
\end{equation}
upon substituting the $\Delta I=1$ transition elements found above and the
vertex amplitudes determined previously. Similarly, we find
\begin{equation}
M_{\psi\pi^0\phi} = \frac{M_{\psi\rho}M_{\rho^0\pi^0\phi}}
{m_\psi^2 - m_\rho^2} + \frac{M_{\phi\rho}M_{\psi\pi^0\rho^0}}
{m_\phi^2-m_\rho^2} + \frac{M_{\pi\eta}M_{\psi\eta\phi}}{m_\pi^2-m_\eta^2}
\simeq 0.0009 {\rm~GeV}^{-1},
\end{equation}
\begin{equation}
M_{\psi\eta\rho^0} = \frac{M_{\psi\rho}M_{\rho^0\eta\rho^0}}
{m_\psi^2 - m_\rho^2} + \frac{M_{\rho\omega}M_{\psi\eta\omega}}
{m_\rho^2-m_\omega^2} + \frac{M_{\eta\pi}M_{\psi\pi^0\rho^0}}
{m_\eta^2-m_\pi^2} \simeq 0.0005 {\rm~GeV}^{-1},
\end{equation}
\begin{equation}
M_{\psi\eta'\rho^0} = \frac{M_{\psi\rho}M_{\rho^0\eta'\rho^0}}
{m_\psi^2 - m_\rho^2} + \frac{M_{\rho\omega}M_{\psi\eta'\omega}}
{m_\rho^2-m_\omega^2} + \frac{M_{\eta'\pi}M_{\psi\pi^0\rho^0}}
{m_\eta'^2-m_\pi^2} \simeq 0.0006 {\rm~GeV}^{-1}.
\end{equation}
The experimental values of these amplitudes, deduced from the measured
decay rates, are, in units of GeV$^{-1}$,
\begin{eqnarray}
|M_{\psi\pi^0\omega}^{exp}| = 0.00067\pm 0.00004,&
|M_{\psi\pi^0\phi}^{exp}| < 0.0009,\nonumber \\
|M_{\psi\eta\rho^0}^{exp}| = 0.00048\pm 0.00003,& 
|M_{\psi\eta'\rho^0}^{exp}| = 0.00040\pm 0.00004.
\end{eqnarray}
All of these answers are in the ``right ball park'' and we venture to
conclude that this $\Delta I=1$, pole-dominated mechanism captures the main
features of such feeble decays.

\section{Some three-body decays}
Here we shall have a look at a few three-body decays, in which pions are
produced in a P-state and are known to be dominated by $\rho$-meson poles;
specifically we shall examine the three processes $\omega \rightarrow 3\pi$
and $\eta ,\eta' \rightarrow 2\pi \gamma$ to see how well they tie up with
the earlier $M$ amplitudes. Before doing so we shall require more accurate
values of the on-shell couplings $g_{\rho\pi\pi}, g_{\omega\rho\pi}$. We
recall that the quark loop alone gives $g_{\rho\pi\pi} = \sqrt{2}g_V
\simeq 5.03$ and $g_{\omega\rho\pi} = 3g_V^2/4\pi^2f_\pi \simeq 10.3$
MeV$^{-1}$. However, in a sigma model, besides the non-strange quark loop
we have to add a meson loop associated with sigma exchange. The effect of
this is to enhance \cite{LSM} the value of $g_{\rho\pi\pi}$ by a factor of
6/5, to about 6.1, which is very close to the experimental value coming from
the $\rho$ decay width. Correspondingly, the $\rho$-$\omega$-$\pi$ coupling
is enhanced \cite{LSM,FN} by a factor of (6/5)$^2$ to $3g_{\rho\pi\pi}^2
/8\pi^2f_\pi \simeq 15$ GeV$^{-1}$.

The amplitude for the $VPPP$ process $\omega\rightarrow 3\pi$ may be written
in the covariant form $A_{\omega\pi_1\pi_2\pi_3} =
\epsilon_{\kappa\lambda\mu\nu}\varepsilon^\kappa p_1^\lambda p_2^\mu p_3^\nu
M_{\omega\pi\pi\pi}$, where the scalar amplitude $M$ is dominated by $\rho$
mesons \cite{GMSW} in each of the three possible two-body channels:
\begin{equation}
M_{\omega\pi\pi\pi} = 2g_{\omega\rho\pi}g_{\rho\pi\pi}\left[\frac{1}{m_\rho^2-  
s} +\frac{1}{m_\rho^2-t} + \frac{1}{m_\rho^2-u}\right]\simeq 1480{\rm ~GeV}^{-3},
\end{equation}
In deriving this result, we have averaged over the Mandelstam variables:
$\langle s \rangle = \langle t \rangle = \langle u \rangle = m_\omega^2/3 +
m_\pi^2 \simeq 0.223$ GeV$^2$. If we then follow Thews' phase space analysis
\cite{Thews}, we get
\begin{equation}
\Gamma(\omega \rightarrow 3\pi)=|M_{\omega\pi\pi\pi}|^2m_\omega^7Y_\omega/
(2\pi)^3
\end{equation}
with the constant matrix element giving $Y_\omega = 4.57\times 10^{-6}.$ In
this way, we predict $\Gamma(\omega\rightarrow 3\pi)\simeq 7.3$ MeV, very
close to the observed rate of $7.5\pm 0.1$ MeV.

Our next object of study, $\eta\rightarrow \pi^+\pi^-\gamma$, is dominated
by a $\rho^0$ meson pole, but just in the two-pion channel \cite{Chanowitz}.
The process also takes the same covariant form as the previous case. Here we
estimate
\begin{equation}
 M_{\eta\pi\pi\gamma} = 2g_{\rho\pi\pi}M_{\eta\rho\gamma}/(m_\rho^2-s)
 \simeq 9.74 {\rm ~GeV}^{-3},
\end{equation}
upon substituting the $\eta$-$\rho$-$\gamma$ amplitude found earlier
(Table I) and going to the soft pion limit. Once more, we find
\begin{equation}
\Gamma(\eta\rightarrow\pi\pi\gamma) = |M_{\eta\pi\pi\gamma}|^2m_\eta^7
Y_\eta/(2\pi)^3 \simeq 55 {\rm~eV},
\end{equation}
for a phase space factor \cite{Thews} of $Y_\eta = 0.98\times 10^{-5}$.
The prediction (31) compares well with the observed rate of $56\pm 5$ eV.

Finally we consider the $\eta'\rightarrow \pi^+\pi^-\gamma$ decay. Unlike
the previous two examples, the $\rho^0$ pole is now well within the physical
region. The latest particle data tables state that $\Gamma(\eta'\rightarrow
2\pi\gamma$, including nonresonant contributions) = $61\pm 5$ keV, whereas
our prediction (see Table I) is that $\Gamma(\eta'\rightarrow\rho\gamma) =
|M_{\eta'\rho\gamma}|^2p_{\rho\gamma}^3/4\pi \simeq 65$ keV. This strongly
suggests that any nonresonant P-wave contact contributions are quite small
and that the $\rho$ poles accurately match the data. Incidentally it also
confirms that the pseudoscalar mixing angle is $\phi_P \simeq 42^o$, rather
than $\phi \sim 35^o$, coupled with the anomaly, as the latter angle gives a
rate for $\eta'\rightarrow\pi\pi\gamma$ which is a factor of 20 less than
the observed rate \cite{VH}.

\section{Conclusions}

Taking stock of the results, we appear to have succeeded in estimating some
34 $VVP$ amplitudes with reasonable accuracy (see the Tables) barring two or
three rare decays involving $\eta'$ and $\omega$ mesons---which are
themselves extracted from one or two experiments that surely deserve
repeating in the future. The success extended to a few abnormal parity three
body decays which are known to be dominated by P-wave pion pairs. We do not
regard the coupling constants $g_P$ and $g_V$ which provide the overall scales 
to be `parameters', because they are predicted in [1] or are
derived from other processes and have well-established magnitudes; 
nor do we think that the quark masses are
variables at our disposal, since they are also fixed by other means. The
mixing angles $\phi_P$ and $\phi_V$ are parameters but they too can be found
through processes other than the ones which we have considered in this paper.
The only (four) variables which we were truly able to adjust were the mixing
angles describing the non-$c\bar{c}$ content of $\eta_c$ and $\psi$.

Perhaps the most surprising aspect of the analysis is the fact that the
amplitudes, which we estimated in the soft limit, have reasonable magnitudes
even for external heavy meson $c\bar{c}$ composites. We really did not
expect this and we think that the conclusions are rather deep and point to
some underlying heavy mass scale, way beyond the usual QCD scale of about
300 MeV. To put this deduction into perspective, consider some
field-theoretic model which includes meson interactions with spinors; assume
it is renormalizable for definiteness. Then the equation for the proper
vertex function $\Gamma(p,q)$ (where $p$ is the meson momentum and $q$ is
the relative fermion momentum) will take the form
$$\Gamma(p,q)=Z\gamma + g^2\int d^4q'\,\,
S(p/2+q')\Gamma(p,q')S(-p/2+q')*K_p(q,q'), $$
where $K$ is the appropriate interaction kernel for the model in question,
$S$ stands for the spinor propagator and $Z\gamma$ is the renormalized
contact term. If one specialises to pseudoscalar mesons and ignores fermion
dressing for simplicity, the equation has the generic scalar form,
$$A(p,q) = Z_p + g^2\int d^4q'\,\,A(p,q')*K_p(q,q')/[p^2/4 - q'^2 - m^2]. $$
The subscript $p$ has been attached to the renormalization constant $Z$ to
remind ourselves that the renormalization procedure is often momentum
dependent, although the infinite part of $Z$ (or a $1/(D-4)$ pole term in
the dimensional method) is of course insensitive to $p$. The limit as
$p\rightarrow 0$ of the equation above is readily taken and describes the
meson vertex in the (off-shell) soft limit.

If the pseudoscalar field is actually a composite of the fermions, then $Z$
must vanish on the meson mass shell ($p^2=M^2$) in order that the vertex
equation reduces to a homogeneous Bethe-Salpeter equation. The results which
we found indicate that the products of vertex functions and quark propagators
running round the loop are rather insensitive to external momentum since
they suggest that an amplitude like $A(p,q)$ above depends very little on
$p$, {\em i.e.} the $q$-dependence on the right hand side has compensating
effects from form factors and propagators. (It is striking that the
propagator denominators would appear to depend sensitively on the binding
energy near the constituent threshold $p^2 = 4m^2$ in the infrared region of
$q'$, and there is no sign of this.) Therefore it indicates that $Z_p$
depends very little on $p$ and that the homogeneous Bethe-Salpeter equation
can very likely be extrapolated to $p=0$ with relatively little change. Our
guess is that such loop integrals are really dominated by the ultraviolet
region and a mass scale $\Lambda$ of at least 10 GeV---much higher than the
standard QCD scale---associated with the ratio $p^2/\Lambda^2$. Indeed, the
seemingly weak dependence of our results on $p$ are the most extraordinary
part of this work and trying to understand the issue properly is an exciting
avenue of future research; our discussion above does scant justice to the
problem.

\acknowledgements
This research was supported by the Australian Research Council. MDS
appreciates the hospitality of the University of Tasmania where a substantial
proportion of the research work was carried out.

\appendix
\section*{Pseudoscalar mixing angle}
Since we are using quark loops to predict $PVV$ amplitudes involving $\eta$
or $\eta'$ pseudoscalar mesons, it is imperative that we determine the most
important $\eta -\eta'$ mixing angle $\phi_P$ from processes other \cite{BS}
than $PVV$ decays. Here we shall consider two other ways of fixing this angle.

First we consider tensor $(T)\rightarrow PP$ decays, $a_2(1320)\rightarrow
K\bar{K}, \quad \eta\pi,\quad \eta'\pi$. Given the respective percentage
branching ratios, from the PDG tables, $4.9\pm 0.8, 14.5\pm 1.2, 0.53\pm
0.09$, and remembering the $p^5$ phase space, we may deduce the two
independent ratios,
\begin{equation}
B(a_2\rightarrow\eta\pi/K\bar{K}) = (p_{\eta\pi}/p_K)^5\cdot 2\cos^2\phi_P =
2.96\pm 0.53
\end{equation}
\begin{equation}
B(a_2\rightarrow\eta'\pi/\eta\pi) = (p_{\eta'\pi}/p_{\eta\pi})^5\tan^2\phi_P
=0.037\pm 0.007.
\end{equation}
Inserting the known magnitudes $(p_{\eta\pi},p_{\eta'\pi},p_K) =
(535,287,437)$ MeV, into this pair of equations, we may conclude from the
first that $\phi_P=43\pm 5^o$, and from the second that $\phi_P = 42\pm 3^o$.
Likewise the measured $K_2^*\rightarrow K\eta/K\pi$ branching ratio of 0.003
requires \cite{BES} $\phi_P = 41\pm 4^o$. As well, there are the tensor
decays $f,f'\rightarrow \pi\pi,\quad K\bar{K},\quad \eta\eta$, but then the
tensor angle $\phi_T$ enters the analysis; yet $\phi_P\sim 42^o$ survives
the results of the analysis.

A second way of extracting $\phi_P$ is through the QCD quark-annihilation
diagram, involving at least two gluon exchange. Diagonalization of the NS-S
meson mass matrix leads \cite{JS} to the $\eta -\eta'$ mixing angle,
\begin{equation}
\phi_P=\arctan\left[\frac{(m^2_{\eta'}-2m^2_K+m^2_\pi)(m_\eta^2-m_\pi^2)}
  {(2m_K^2-m_\pi^2-m_\eta^2)(m^2_{\eta'}-m^2_\pi)}\right]^{1/2} = 41.9^o.
\end{equation}
The close agreement between $\phi_P$ determined by $T\rightarrow PP$ data
and QCD theory justifies our use of $\phi_P \simeq 42^o$ in Tables I and II.

In the singlet-octet basis, the mixing angle $\theta$ is instead given by
\cite{BES,JS}
$$ \theta = \phi - \arctan\sqrt{2} = \phi - 54.7^0 ,\quad {\rm or}$$
$$\cos\phi =\frac{1}{\sqrt{3}}\cos\theta - \sqrt{\frac{2}{3}}\sin\theta,
\qquad\sin\phi =\sqrt{\frac{2}{3}}\cos\theta +\frac{1}{\sqrt{3}}\sin\theta.$$
Thus for the pseudoscalar mesons, $\theta_P \simeq -13^0$, which is midway
between the original Gell-Mann-Okubo value of $-10^0$ and most recent
determinations of $-20^0$. In this connection, it is interesting to compare
the quark triangle predictions,
\begin{mathletters}
\begin{equation}
|M_{\eta\gamma\gamma}| = \frac{\alpha}{3\sqrt{3}\pi f_\pi}\left[
   (5-\frac{2\hat{m}}{m_s})\cos\theta_P -
   \sqrt{2}(5+\frac{\hat{m}}{m_s})\sin\theta_P\right],
\end{equation}
\begin{equation}
|M_{\eta'\gamma\gamma}|=\frac{\alpha}{3\sqrt{3}\pi f_\pi}\left[
  (5-\frac{2\hat{m}}{m_s})\sin\theta_P +
  \sqrt{2}(5+\frac{\hat{m}}{m_s})\cos\theta_P\right],
\end{equation}
\end{mathletters}
with the predictions obtained via the AVV anomaly \cite{VH},
\begin{mathletters}
\begin{equation}
|M^{anom}_{\eta\gamma\gamma}|=\frac{\alpha}{\sqrt{3}\pi f_\pi}\left[
    \frac{f_\pi}{f_8}\cos\theta_P -
    2\sqrt{2}\frac{f_\pi}{f_0}\sin\theta_P\right],
\end{equation}
\begin{equation}
|M^{anom}_{\eta'\gamma\gamma}|=\frac{\alpha}{\sqrt{3}\pi f_\pi}\left[
    \frac{f_\pi}{f_8}\sin\theta_P +
    2\sqrt{2}\frac{f_\pi}{f_0}\cos\theta_P\right].
\end{equation}
\end{mathletters}
Comparing (A.4) with (A.5), we see that
\begin{mathletters}
\begin{equation}
f_8/f_\pi = 3/(5-2\hat{m}/m_s) \simeq 0.83,
\end{equation}
\begin{equation}
f_0/f_\pi = 6/(5+\hat{m}/m_s) \simeq 1.05.
\end{equation}
\end{mathletters}
Both (A.6ab) exhibit U(3) symmetry in the limit $m_s\rightarrow \hat{m}$.
However for the actual constituent quark mass ratio (obtained from the
GTR or magnetic moments) of $m_s/\hat{m} = (2f_K/f_\pi-1) \simeq 1.44$,
the numerical values in (A.6) follow. Note that the anomaly version\cite{VH}
estimates $f_8/f_\pi\simeq 1.3$ via chiral perturbation theory but also
finds $f_0/f_\pi\simeq 1.04$, which is close to our prediction; we agree
with their justification that $f_0/f_\pi$ being near unity is what one
expects if the singlet state and the pion have the same wave function.

\begin{table}
\caption{Dominant strong and radiative decays that are somewhat insensitive to 
mixing angles. Under the column of theoretical formulae, $M_{PVV}$, we use the abbreviations,
$A\!=\!e^2/4\pi^2f_\pi\!=\!\alpha/\pi f_\pi\simeq 0.025$, $B\!=\! eg_Vg_P/8\pi^2\hat{m}\simeq 0.207$, $C\!=\! g_V^2/4\pi^2f_\pi\!=\! \sqrt{2}g_V^2g_P/8\pi^2\hat{m}\simeq 3.47$,
in units of GeV$^{-1}$, and $r_q = \hat{m}/m_q$ for the constituent quark mass 
ratio. The latter are determined by the assumed values (in GeV) $\hat{m} = 0.34, 
m_s = 0.49, m_c = 1.5, m_b = 4.8$}

\begin{tabular}{ccccc}
Process&$\Gamma^{exp}$(MeV)&$|M^{exp}|$ (GeV$^{-1}$) & $M_{PVV}$
& $|M^{thy}|$ (GeV$^{-1}$)\\
\tableline
$\pi^0\rightarrow\gamma\gamma$ & $(7.74\pm 0.52)\times 10^{-6}$ & $0.0251\pm
0.008$ & $A$ & 0.0250\\
$\eta\rightarrow\gamma\gamma$ & $(4.6\pm 0.4)\times 10^{-4}$ & $0.024\pm
0.001$ & $A[5\cos\phi_P - \sqrt{2}r_s\sin\phi_P]/3$ & 0.0255\\
$\eta'\rightarrow\gamma\gamma$ & $(4.3\pm 0.4)\times 10^{-3}$ & $0.032\pm
0.001$ & $A[5\sin\phi_P + \sqrt{2}r_s\cos\phi_P]/3$ & 0.0335\\
$\eta_c\rightarrow\gamma\gamma$ & $(7.5\pm 2.5)\times 10^{-3}$ & $0.0075\pm
0.0013$ & $4\sqrt{2}Ar_c/3$ & 0.0107\\
\tableline
$\eta'\rightarrow\rho\gamma$ & $0.061\pm 0.005$ & $0.40\pm 0.02$ &
$3B\sin\phi_P$ & 0.42\\
$\eta'\rightarrow\omega\gamma$ & $(6.1\pm 0.8)\times 10^{-3}$ & $0.14\pm
0.01$ & $B[\cos\!\phi_V\sin\!\phi_P\!+\!2r_s\!\sin\!\phi_V\cos\!\phi_P]$ &
0.15\\
\tableline
$\rho^{\pm}\rightarrow\pi^{\pm}\gamma$ & $0.068\pm 0.007$ & $0.22\pm 0.01$ &
$B$ & 0.21\\
$\rho^0\rightarrow\eta\gamma$ & $0.036\pm 0.002$ & $0.45\pm 0.07$ &
$3B\cos\phi_P$ & 0.46\\
$\omega\rightarrow\pi^0\gamma$ & $0.71\pm 0.04$ & $0.70\pm 0.02$ &
$3B\cos\phi_V$ & 0.62\\
$\omega\rightarrow\eta\gamma$ & $(5.5\pm 1)\times 10^{-3}$ & $0.16\pm 0.01$ &
 $B[\cos\phi_P\cos\phi_V\!-\!2r_s\sin\phi_P\sin\phi_V]$ & 0.14\\
$\phi\rightarrow\pi^0\gamma$ & $(5.8\pm 0.5)\times 10^{-3}$ & $0.042\pm
0.002$ & $3B\sin\phi_V$ & 0.04\\
$\phi\rightarrow\eta\gamma$ & $0.056\pm 0.002$ & $0.21\pm 0.01$ &
$B[2r_s\sin\phi_P\cos\phi_V\!+\!\sin\phi_V\cos\phi_P]$ & 0.20\\
$\phi\rightarrow\eta'\gamma$ & $(5\pm 3)\times 10^{-4}$ & $0.30\pm 0.08$ &
$B[\sin\phi_V\sin\phi_P\!-\! 2r_s\cos\phi_V\cos\phi_P]$ & 0.21\\
$\phi\rightarrow\pi^0\rho^0$ & $<0.23\pm 0.01$ & $<1.18\pm 0.02$ &
$3C\sin\phi_V$ & 0.69\\
$\psi\rightarrow\eta_c\gamma$ & $(1.1\pm 0.4)\times 10^{-3}$ & $0.17\pm
0.03$ & $4Br_c$ & 0.19\\
$K^{*+}\rightarrow K^+\gamma$ & $0.050\pm 0.006$ & $0.252\pm 0.015$ &
$\frac{Br_s[2J(r_s)-J(1/r_s)]}{(1+r_s)}$ & 0.20\\
$K^{*0}\rightarrow K^0\gamma$ & $0.116\pm 0.011$ & $0.39\pm 0.02$ & 
$-\frac{Br_s[J(r_s)+J(1/r_s)]}{(1+r_s)}$ & 0.34\\
$D^{*+}\rightarrow D^+\gamma$ & $<0.0014$ & $<0.14\pm 0.10$ &
$\frac{Br_c[-J(r_c)+2J(1/r_c)]}{(1+r_c)}$ & 0.03\\
$D^{*0}\rightarrow D^0\gamma$ & $<0.8$ & $<3.4$ &
$\frac{2Br_c[J(r_c)+J(1/r_c)]}{(1+r_c)}$ & 0.26\\
$D^{*\pm}_s\rightarrow D_s^\pm\gamma$ & $<1.8$ & $<5.0$ &
$\frac{Br_sr_c[J(r_c/r_s)+2J(r_s/r_c)]}{(r_s+r_c)}$ & 0.04
\end{tabular}
\label{Table1}
\end{table}

\begin{table}
\caption{Rare decays which are sensitive to the four mixing parameters
$\delta$ and $\epsilon$ in (14) and (15). $A$, $B$ and $C$ are the same
abbreviations as in Table I, and again $r_q = \hat{m}/m_q$ stands for the
constituent quark mass ratio. In writing down the $\psi$ amplitudes, we have
set $\cos\phi_V=1$ as a first approximation and ignored terms of order
$\epsilon\delta$, which can be justified a posteriori.}

\begin{tabular}{ccccc}
Process&$\Gamma^{exp}$(keV)&$|M^{exp}|$ (TeV$^{-1}$) & $M_{PVV}$
& $|M^{thy}|$ (TeV$^{-1}$)\\
\tableline
$\eta_c\rightarrow\rho^0\rho^0$ & $(113\pm 40)$ & $37\pm 6$ & 
$-3C(\delta_c\cos\phi_P+\delta'_c\sin\phi_P)$ & 37\\
$\eta_c\rightarrow\omega\omega$ & $<84$ & $<32$ & $-3C(\delta_c\cos\phi_P +
\delta'_c\sin\phi_P)$ & 37\\
$\eta_c\rightarrow\phi\phi$ & $94\pm 30$ & $43\pm 6$ &
$3\sqrt{2}C(\delta_c\sin\phi_P-\delta'_c\cos\phi_P)r_s$ & 43\\
$\eta_c\rightarrow K^{*0}\bar{K}^{*0}$ & $58\pm 30$ & $17\pm 4$ &
$\frac{3Cr_s}{1+r_s}\left[\begin{array}{c} - (\delta_c\cos\phi_P+
\delta'_c\sin\phi_P)J(r_s)\\
 +\frac{r_s}{\sqrt{2}}(\delta_c\sin\phi_P-\delta'_c\cos\phi_P)J(1/r_s)
 \end{array}\right]$ & 16\\
\tableline
$\psi\rightarrow\pi^0\gamma$ & $0.003\pm 0.001$ & $0.19\pm 0.02$ &
$-3B\epsilon_c$ & 0.13\\
$\psi\rightarrow\eta\gamma$ & $0.075\pm 0.008$ & $0.92\pm 0.05$ &
$B[-\epsilon_c\cos\phi_P-2r_s\epsilon'_c\sin\phi_P+4r_c\delta_c]$ & 1.1\\
$\psi\rightarrow\eta'\gamma$ & $0.37\pm 0.04$ & $2.3 \pm 0.1$ &
$B[-\epsilon_c\sin\phi_P+2r_s\epsilon'_c\cos\phi_P+4r_c\delta'_c]$ & 0.1\\
$\psi\rightarrow\pi^0\rho^0$ & $0.37\pm 0.02$ & $2.1\pm 0.1$ &
$-3C\epsilon_c$ & 2.1\\
$\psi\rightarrow\eta\omega$ & $0.14\pm 0.02$ & $1.4\pm 0.1$ &
$-3C[\epsilon_c\cos\phi_P+\sqrt{2}r_s\epsilon'_c\sin\phi_V\sin\phi_P]$&1.6\\
$\psi\rightarrow\eta\phi$ & $0.057\pm 0.008$ & $0.97\pm 0.07$ &
$-3C[\epsilon_c\cos\phi_P\sin\phi_V-\sqrt{2}r_s\epsilon'_c\sin\phi_P]$&0.62\\
$\psi\rightarrow\eta'\omega$ & $0.015\pm 0.002$ & $0.52\pm 0.04$ &
$-3C[\epsilon_c\sin\phi_P-\sqrt{2}r_s\epsilon'_c\sin\phi_V\cos\phi_P]$&1.4\\
$\psi\rightarrow\eta'\phi$ & $0.029\pm 0.003$ & $0.80\pm 0.04$ &
$-3C[\epsilon_c\sin\phi_P\sin\phi_V+\sqrt{2}r_s\epsilon'_c\cos\phi_P]$&0.68\\
$\psi\rightarrow K^0\bar{K}^{*0}$ & $0.37\pm 0.03$ & $2.3\pm 0.1$ &
$-\frac{6Cr_s[\epsilon_cJ(r_s)+\epsilon'_cJ(1/r_s)]} {(1+r_s)}$ & 2.2\\
$\psi\rightarrow K^+\bar{K}^{*-}$ & $0.43\pm 0.07$ & $2.5\pm 0.2$ &
$-\frac{6Cr_s[\epsilon_cJ(r_s)+\epsilon'_cJ(1/r_s)]} {(1+r_s)}$ & 2.2
\end{tabular}
\label{Table2}
\end{table}

\newpage
\begin{figure}[b!] 
\centerline{\epsfig{file=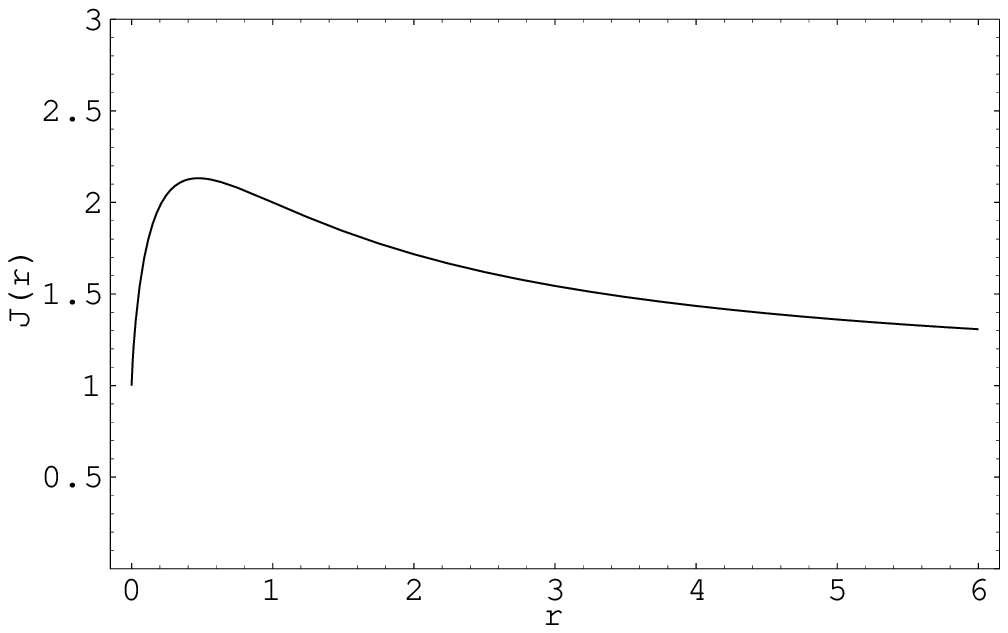}}
\vspace{1pt}
\caption{Plot of the function $J(r)$ from $r=0$ to 6.}
\label{fig1}
\end{figure}


\begin{references}

\bibitem[*]{A1} E-mail: Bob.Delbourgo@utas.edu.au
\bibitem[\dag]{A2} E-Mail: D.Liu@utas.edu.au
\bibitem[\ddag]{A3} Permanent address: Physics Department, University of
Arizona, Tucson, AZ 85721; E-Mail: scadron@physics.arizona.edu;
\bibitem{LSM} R.~Delbourgo and M.D.~Scadron, Mod. Phys. Lett. {\bf A10}, 251
(1995); A.~Bramon, Riazzuddin and M.D.~Scadron, J. Phys. {\bf G24}, 1 (1998).
These are archived in hep-ph/9807505 and hep-ph/9709274 respectively.
\bibitem{CA} S.L.~Adler, Phys. Rev. {\bf 177}, 2426 (1969); J.S.~Bell and
R.~Jackiw, {Nuovo Cim.} {\bf 60}, 47 (1969).
\bibitem{PDG} Particle Data Group, C. Caso et al, Eur. Phys. J. {\bf C3}, 1
(1998). All the data in our tables refers to the average values appearing in
this reference; these averages are obtained from many, many experiments.
\bibitem{Steinberger} J.~Steinberger, Phys. Rev. {\bf 76}, 1180 (1949).
\bibitem{BWL} B.W.~Lee, {\em Chiral Dynamics} (Gordon and Breach, NY, 1972)
p. 12.
\bibitem{GML} M.~Gell-Mann and M.~Levy, Nuovo Cim. {\bf 16}, 705 (1960).
\bibitem{1996} R.M.~Barnett et al, Phys. Rev. {D54}, 1 (1996).
\bibitem{CB} S.A.~Coon and R.C.~Barrett, Phys. Rev. {\bf C36}, 2189 (1987);
S.A.~Coon, P.C.~McNamee and M.D.~Scadron, Nucl Phys. {\bf A249}, 483 (1975);
{\em ibid} {\bf A287}, 381 (1977).
\bibitem{DS} R.~Delbourgo and M.D.~Scadron, Int. J. Mod. Phys. {\bf 13A},
657 (1998).
\bibitem{CS} S.A.~Coon and M.D.~Scadron, Phys. Rev. {\bf C51}, 2923 (1995).
\bibitem{FN} P.G.O.~Freund and S.~Nandi, Phys. Rev. Lett. {\bf 32}, 181
(1974); S.~Rudaz, Phys. Lett. {\bf B145}, 281 (1984).
\bibitem{GMSW} M.~Gell-Mann, D.~Sharp and W.~Wagner, Phys. Rev. Lett.
{\bf 8}, 261 (1962).
\bibitem{Thews} R.L.~Thews, Phys. Rev. {\bf D10}, 2993 (1974); see his
appendix.
\bibitem{Chanowitz} M.S.~Chanowitz, Phys. Rev. Lett. {\bf 35}, 977 (1975);
{\em ibid} {\bf 44}, 59 (1980); M.~Benayoun et al, Z. Phys. {\bf C58}, 31
(1993); {\em ibid} {\bf C65}, 399 (1995).
\bibitem{VH} E.P.~Venugopal and B.R.~Holstein, Phys. Rev. {\bf D57}, 4397
(1998).
\bibitem{BS} A.~Bramon and M.D.~Scadron, Phys. Lett. {\bf B234}, 346 (1990);
Phys. Rev. {\bf D40}, 3779 (1989).
\bibitem{BES} A.~Bramon, R.~Escribano and M.D.~Scadron, Eur. Phys. J
{\bf C}, in press; see also hep-ph/9711229.
\bibitem{JS} H.F.~Jones and M.D.~Scadron, Nucl. Phys. {\bf B155}, 409 (1979);
M.D.~Scadron, Phys. Rev. {\bf D29}, 2076 (1984).
\end{references}
\end{document}